\documentclass[aps,draft]{revtex4}
\usepackage[dvips]{graphicx}
\begin{document}

\title{Relativistic Thermodynamics of Magnetized Fermi Electron Gas}

\author{ Nodar L. Tsintsadze and Levan N. Tsintsadze}
\affiliation{ Faculty of Exact and Natural Sciences, Andronikashvili Institute of Physics,  Tbilisi State University, Tbilisi, Georgia }

\date{\today}

\begin{abstract}

To study the relativistic thermodynamic properties of a Fermi gas in a strong magnetic field, we construct the relativistic thermodynamic potential by the relativistic Fermi distribution function taking into account that the motion of particles in a plane perpendicular to the magnetic field is quantized. With this general potential at hand, we investigate all the thermodynamic quantities as a function of densities, temperatures and the magnetic field. We obtain a novel set of adiabatic equations. Having the expression of the pressure and adiabatic state equations, we determine the sound velocity for several cases revealing a new type of sound velocity. Finally, we disclose the magnetic cooling in the quantized electron Fermi gas, which is based on an adiabatic magnetization in contrast to the known adiabatic demagnetization.
\end{abstract}


\maketitle

\section{INTRODUCTION}

The influence of strong or superstrong magnetic field on the thermodynamic properties of medium and the propagation of proper waves is an important issue in supernovae and neutron stars, the convective zone of the sun, the early prestellar period of the evolution of the universe, as well as in the contemporary problems of laser-matter interaction. Ginsburg was the first to suggest that a collapse of star must be accompanied by the generation of superstrong magnetic field \cite{gins}. Based on the astrophysical data, the surface magnetic field of a neutron star
is $H\sim 10^{11}-10^{13}G$, and the internal field can reach $H\sim 10^{15}G$ or
even higher \cite{lan}-\cite{lip}. As was shown by Bisnovati-Kogan \cite{bis}, the presence of rotation of stars may increase the magnetic field by an additional factor of $10^3-10^4$. In such strong magnetic fields, it is expected that the thermodynamic properties and wave dynamics in degenerate electron gas would be quite different governed by the quantum effects. This is true when the characteristic energy of electron on a Landau level reaches the nonrelativistic limit of the electron
chemical potential $\mu=\varepsilon_F=\frac{\hbar\mid e\mid H}{2m_ec},$ i.e.
$H=H_S\frac{v_F^2}{c^2}$, where $H_S=\frac{m_e^2c^3}{\mid e\mid \hbar}=
4.4\cdot 10^{13}G$ is the Schwinger's magnetic field, $v_F=p_F/m_e=(3\pi^2)^{1/3}\hbar n_e^{1/3}/m_e$ is the speed of electrons at the Fermi surface, $\hbar$ is the Planck constant divided by $2\pi$, $m_e$ is the electron rest mass, c is the speed of light in vacuum, $\mid e\mid$ is the magnitude of the electron charge and $n_e$ is the density of electrons.

In the past we have developed the relativistic thermodynamics of electron-ion or electron-positron-ion plasmas for arbitrary temperatures in the presence of electromagnetic radiation \cite{ltsin95},\cite{ltsin96} and studied numerically in Ref.\cite{ltsin97}. Whereas the relativistic statistical thermodynamics of dense photon gas we have formulated recently \cite{ltsin07}. In the present paper, we create the relativistic thermodynamics of magnetized quantum electron gas, which is becoming of increasing current interest motivated by its potential application in modern technology, e.g. metallic and semiconductor nanostructures - such as metallic nanoparticles, metal clusters, thin metal films, spintronics, nanotubes, quantum well and quantum dots, nano-plasmonic devices, quantum x-ray free-electron lasers, etc. Moreover, quantum electron gas is common in planetary interiors, in compact astrophysical objects (e.g., the interior of white dwarf stars, magnetospheres of neutron stars and magnetars, etc.).

Properties of linear electron oscillations in a dense Fermi plasma have been studied in the remote past \cite{gol}-\cite{bp}. Despite extensive theoretical efforts (for recent review see \cite{pad}) since then, there are questions and issues which has to be clarified. Answers to some salient questions are given in Refs.\cite{tsin},\cite{nodar}, with a new type of quantum kinetic equations of the Fermi particles of various species and a general set of fluid equations. This kinetic
equation for the Fermi electron gas was used to study the propagation of small longitudinal perturbations, deriving a quantum dispersion equation. Later the dispersion properties of linear oscillations of quantum electron-ion \cite{tsin10},\cite{ell} and electron-positron-ion \cite{tsin12} plasmas, as well as of neutral $H_e^3$ \cite{tsin11} have been studied. The effects of the quantization of the orbital motion of electrons and the spin of electrons on the propagation of longitudinal waves in the Fermi gas have been also reported recently \cite{ltsin}. As is well known, the strong magnetic field in the Fermion gas leads to two magnetic effects \cite{land}. Namely, they are the Pauli paramagnetism due to the spin of electrons and the Landau diamagnetism due to the quantization of the orbital motion of electrons. In Ref.\cite{ltsin}, a novel dispersion relation of the longitudinal wave propagating along a magnetic field was derived, which exhibits the strong dependence on the magnetic field in radical contrast to the classical case. Note that in the classical limit the propagation of longitudinal wave along the magnetic field is unaffected by the magnetic field.

We also note that the diamagnetic effect has a purely quantum nature and in the classical electron gas it is absent,  because in a magnetic field the Lorentz's force $\frac{e}{c}\vec{v}\times\vec{H}$ acts on a particle in the perpendicular direction to a velocity $\vec{v}$, so that it cannot produce work on the particle. Hence, its energy does not depend on the magnetic field. However, as was shown by Landau, the situation radically changes in the quantum mechanical theory of magnetism. The point is that in a constant magnetic field the electrons, under the action of it, rotate in circular orbits in a plane perpendicular to the field $\vec{H}_0(0,0,H_0)$. Therefore, the motion of the electrons can be resolved into two parts: one along the magnetic field, in which the longitudinal component of energy is not changed and is infinite, i.e. the energy of particle is not quantized, $E_\parallel=\frac{p_z^2}{2m_e}$, and the second in a plane perpendicular to $H_0$ (the transverse component), in which the electron rotates in circular orbits with the cyclotron frequency and is finite, i.e. the transverse component of energy is quantized. Because the transverse motion of electrons is similar to the motion of linear harmonic oscillator, the corresponding energy spectrum consists of discrete energy levels \cite{landS},\cite{ber}
\begin{eqnarray*}
E_\perp=(\ell+\frac{1}{2})\hbar\omega_c \ ,
\end{eqnarray*}
where $\ell$ is the orbital quantum number ($\ell=0,1,2,...$) and $\omega_c=\frac{\mid e\mid H}{m_ec}$ is the cyclotron frequency of the electron.

Thus, in the nonrelativistic case the net energy of electron in a magnetic field without taking into account its spin is
\begin{eqnarray*}
E(\ell, p_z)=(\ell+\frac{1}{2})\hbar\omega_c+\frac{p_z^2}{2m_e} \ .
\end{eqnarray*}
If the particle has a spin, the intrinsic magnetic moment of the particle interacts directly with the magnetic field. The correct expression for the energy is obtained by adding an extra term $\vec{\mu}\vec{H}$ corresponding to the energy of the magnetic moment $\vec{\mu}$ in the field $\vec{H}$. Hence, the electron energy levels $\varepsilon_e^{\ell,\sigma}$ are determined in the nonrelativistic limit by the expression
\begin{eqnarray}
\label{eel}
\varepsilon_e^{\ell,\sigma}=\frac{p_z^2}{2m_e}+(2\ell+1+\sigma)\beta_B H\ ,
\end{eqnarray}
where $\sigma$ is the operator the z component of which describes the spin orientation $\vec{s}=\frac{1}{2}\vec{\sigma }$ ($\sigma=\pm 1$), and $\beta_B=\frac{\mid e\mid\hbar}{2m_ec}$ is the Bohr magneton.

Whereas the relativistic expression of the energy reads
\begin{eqnarray}
\label{rel}
\varepsilon_e^{\ell,\sigma}=m_ec^2\sqrt{1+(2\ell+1+\sigma)\frac{\hbar\omega_c}{m_ec^2}+\frac{p_z^2}{m_e^2c^2}}\ .
\end{eqnarray}
From the expression (\ref{rel}) one sees that the energy spectrum of electrons consist of the lowest Landau level, $\ell=0$, $\sigma=-1$, and pairs of degenerate levels with opposite polarization, $\sigma=1$. Thus each value with $\ell\neq 0$ occurs twice, and that with $\ell=0$ once. Therefore, in the nonrelativistic limit $\varepsilon_e^{\ell,\sigma}$ can be rewritten as
\begin{eqnarray}
\label{reel}
\varepsilon_e^{\ell,\sigma}=\varepsilon_e^{\ell}=\frac{p_z^2}{2m_e}+\ell\hbar\omega_c \ ,
\end{eqnarray}
and the relativistic expression is
\begin{eqnarray}
\label{rreel}
\varepsilon_e^{\ell}= m_ec^2\sqrt{1+2\frac{\hbar\omega_c}{m_ec^2}\ \ell+\frac{p_z^2}{m_e^2c^2}}\ .
\end{eqnarray}

The number of quantum states $\frac{2V\mid e\mid Hdp_z}{(2\pi\hbar)^2c}$ of a particle moving in a volume V and the interval $dp_z$ for any value of $\ell$ in a new variables $u=\frac{p_z}{m_e c}$, $\overline{\varepsilon}_e^{\ell}=\frac{\varepsilon_e^{\ell}}{m_e c^2}=\sqrt{1+D\ell+u^2}$, $D=2\frac{\hbar\omega_c}{m_e c^2}$, $\overline{\mu}=\frac{\mu}{m_e c^2}=\sqrt{1+2\frac{\varepsilon_F}{m_e c^2}}$, is
\begin{eqnarray}
\label{nqs}
\frac{2V\mid e\mid H dp_z}{(2\pi\hbar)^2c}=\frac{V D du}{2\pi^2\lambda_c^3}\ ,
\end{eqnarray}
where $\lambda_c=\frac{\hbar}{m_e c}$ is the Compton wavelength.

\section{Thermodynamics of Fermi Gas}

The study of thermodynamics of a Fermi gas of sufficiently low temperature and in a strong magnetic field is of fundamental significance. As the gas is compressed, the mean energy of electrons increases and when it becomes comparable or larger than the rest energy of the electrons, the relativistic effects become important. In this case the temperature can be $10^3$-$10^9\ ^{o}K$, and the relativistic degenerate electron gas will be formed. For instance, a supernova explosion happens when the inner part of a star first collapses to very large densities $\rho\approx 10^{12}-10^{14} g/cm^3$. In this case the Fermi energy (without a magnetic field) is $\varepsilon_F\simeq cp_F=(3\pi^2)^{1/3}\hbar cn_e^{1/3}\simeq 10^8 eV$, which is much more than the thermal energy, even for the temperature $T\sim 10^9-10^{10}\ ^{o}K$.

We now discuss in details the density of electrons in the presence of a strong magnetic field. The equilibrium total density of particles is defined as
\begin{eqnarray}
\label{tnd}
n_e=\frac{D}{4\pi^2\lambda_c^3}\sum_{\ell=0}^\infty\int_{-\infty}^\infty du \ f_e(u,\ell)
\end{eqnarray}
with the Fermi distribution function
\begin{eqnarray}
\label{fdf}
f_e(u,\ell)=\frac{1}{\exp \left\{\frac{\overline{\varepsilon}_e^{\ell}-\overline{\mu}}{\Theta}\right\}+1}\ ,
\end{eqnarray}
where $\Theta=\frac{K_BT}{m_e c^2}$ and $K_B$ is Boltzmann's constant.

To evaluate the density $n_e$ from the expression (\ref{tnd}), we shall consider a Fermi gas at the temperature limit
$\mid \ell\hbar\omega_c-\mu\mid\gg T$. In this case the Fermi distribution function is in a good approximation described by the Heaviside step function $H(\mu-\varepsilon_e^\ell)$.

Let us recall some purely quantum mechanical features of a macroscopic system. It is well known that there is an extremely
high density of levels in the energy eigenvalue spectrum of a macroscopic system. We know also that the number of levels in a
given finite range of the energy spectrum of a macroscopic system increases exponentially with the number of particles N in the
system, and separations between levels are given by numbers of the $10^{-N}$. We can therefore conclude that in such case the
spectrum is almost continuous, and a quasi-classical approximation is applicable. Thus, we can replace the summation in Eq.(\ref{tnd}) by an integration $(\sum_1^{\ell_{max}}\rightarrow\int_1^{\ell_{max}}d\ell)$ to obtain after a simple integration an expression of the density $n_e$. The result is
\begin{eqnarray}
\label{fden}
n_e=\frac{D}{2\pi^2\lambda_c^3}\left\{\sqrt{\overline{\mu}^2-1}+\frac{2}{3 D}(\overline{\mu}^2-1-D)^{3/2}\right\}\ .
\end{eqnarray}
In equation (\ref{fden}) the first term is the contribution from the lowest Landau level ($\ell=0$), i.e. this term is associated with the Pauli paramagnetism and self-energy of particles. The second one results from the summation over all higher Landau levels.

If the magnetic field is absent, then we recover the well known expression \cite{landS}
\begin{eqnarray}
\label{wde}
n_e=\frac{1}{3\pi^2\lambda_c^3}(\overline{\mu}^2-1)^{3/2}\ ,
\end{eqnarray}
whereas, if the magnetic field is very strong $D>(\overline{\mu}^2-1)$, the sum in Eq.(\ref{tnd}) vanishes and all electrons are at the main level, which means that the gas is fully polarized and spins of all particles are aligned opposite to the magnetic field. Hence, in this case for the density we have
\begin{eqnarray}
\label{smd}
n_e=\frac{D}{2\pi^2\lambda_c^3}\sqrt{\overline{\mu}^2-1}\ ,
\end{eqnarray}
which in the nonrelativistic limit yields \cite{ltsin}
\begin{eqnarray}
\label{nsmd}
n_e=\frac{\hbar\omega_c m_e p_F}{\pi^2\hbar^3}\ .
\end{eqnarray}
In the ultrarelativistic limit ($\overline{\mu}\gg 1$) from Eq.(\ref{fden}) we obtain
\begin{eqnarray}
\label{rfden}
n_e=\frac{\hbar\omega_cm_e}{\pi^2\hbar^3}\ p_F\left\{1+\frac{1}{3}\frac{
p_F^2}{\hbar\omega_cm_e}\Bigl(1-\frac{2\hbar\omega_cm_e}
{p_F^2}\Bigr)^{3/2}\right\}\ .
\end{eqnarray}
From here it is clear that if $(\overline{\mu}^2-1)<D$, then we get the same in form expression as Eq.(\ref{nsmd}) with the difference that now $p_F>m_e c$.
If suppose that the density of electrons is constant, then from Eq.(\ref{nsmd}) follows an important statement that when the magnetic field increases the momentum along it decreases and a pancake configuration of Fermi energy becomes thin, and at $H\rightarrow\infty$, $p_F$ goes to zero, and in the momentum space we obtain just flat-flatten.

We now write the thermodynamic potential of relativistic electron gas with an arbitrary magnetic field and temperature
\begin{eqnarray}
\label{tpot}
\Omega=-2\frac{V\mid e\mid H}{(2\pi\hbar)^2c}K_BT\sum_{\ell=0}^\infty\int_{-\infty}^\infty dp_z\ln
\Bigl(1+e^{\frac{\mu-\varepsilon_e^\ell}{K_B T}}\Bigr) \ .
\end{eqnarray}
Clearly the evolution of Eq.(\ref{tpot}) is nontrivial, but it is possible to restrict ourselves to the most simple limiting cases. So that we consider the orders of magnitude of characteristic energy $\varepsilon_F$ (the Fermi energy of the highest occupied state), T and $\hbar\omega_c$ of the system.

We shall show that at some condition the thermodynamic potential contains a part which oscillates with a large amplitude as a function of H. In the remote past the oscillations of the magnetic susceptibility of metal single crystals at low temperatures were qualitatively predicted by Landau and experimentally detected by de Haas-van Alphen \cite{landS},\cite{sho}. For the oscillations a necessary condition is $\mu\gg\hbar\omega_c\gtrsim K_BT$. In which case all thermodynamic quantities must oscillate. It should be emphasized that at $\mu\gg\hbar\omega_c$, the dependence of the not oscillatory part of thermodynamic potential on the magnetic field is very weak and the influence of the magnetic filed on all thermodynamic quantities can be ignored.

In order to separate the oscillatory parts of $\Omega$ in Eq.(\ref{tpot}), we use the Poisson formula
\begin{eqnarray}
\label{pois}
\Omega=\sum_{\ell=0}^\infty \Psi(\ell)=\frac{1}{2}\Psi(0)+\int_{1}^\infty d\ell\ \Psi(\ell)+2Re
\sum_{k=1}^\infty\int_0^\infty d\ell\ \Psi(\ell)e^{2\imath\pi k\ell}\ ,
\end{eqnarray}
where
\begin{eqnarray*}
\Psi(\ell)=-2\frac{V\mid e\mid H}{(2\pi\hbar)^2c}\int_{-\infty}^\infty dp_z\ln
\Bigl(1+e^{\frac{\mu-\varepsilon_e^\ell}{T}}\Bigr) \ .
\end{eqnarray*}
In the expression (\ref{pois}) the first two terms describe the monotonous dependence of the thermodynamic potential on the magnetic field and one should take into account the magnetic field if $\hbar\omega_c\sim\mu$, whereas the last terms depict the oscillatory parts, which become essential when $\mu\gg\hbar\omega_c\gtrsim K_BT$. We can, therefore, neglect the last terms in (\ref{pois}) at $\mu\sim\hbar\omega_c$ and calculate only the monotony parts.

We consider the case, when the thermal energy $K_BT$ is much less than the Fermi energy $\varepsilon_F$. In this case the distribution function is appreciably different from unity or zero only in a narrow range of values of the energy $\varepsilon_e^\ell$ close to the limiting energy $\mu=\varepsilon_F$. The width of this transition zone of the Fermi distribution is of the order of $K_BT$. Thus, in this case the logarithm of the grand partition function (\ref{tpot}) can be rewritten in the form
\begin{eqnarray}
\label{grp}
\Omega=-2\frac{V\mid e\mid H}{(2\pi\hbar)^2c}\sum_{\ell=0}^\infty\int_{-\infty}^\infty dp_z\
p_z\ \frac{d\varepsilon_e^\ell}{dp_z}\left\{H(\mu-\varepsilon_e^\ell)+\Bigl(\frac{1}{e^{\frac{
\varepsilon_e^\ell-\mu}{K_B T}}+1}-H(\mu-\varepsilon_e^\ell)\Bigr)\right\}\ .
\end{eqnarray}
The first integral denotes the value of $\Omega_0$ at absolute zero temperature and is the first term in the expansion of the corresponding quantity in powers of the small ratio $T/T_F$, where $T_F=\frac{\varepsilon_F}{K_B}$ is the Fermi degeneracy temperature.

The second term in (\ref{grp}) describes the temperature and magnetic field dependence of thermodynamical potential, including the first order ($T/T_F$) temperature correction. After integration over momentum we obtain
\begin{eqnarray}
\label{ain}
&&\Omega=\Omega_0+\Omega_T=-\frac{V m_ec^2 D}{2\pi^2\lambda_c^3}\times \\ \nonumber &&\left\{\frac{1}{2}(\overline{\mu}\sqrt{\overline{\mu}^2-1}-\ln(
\overline{\mu}+\sqrt{\overline{\mu}^2-1})+
\frac{1}{2}\sum_{\ell=1}^{\ell_{max}}\Bigl(
\overline{\mu}\sqrt{\overline{\mu}^2-1-D\ell}-
(1+D\ell)\ln\frac{\overline{\mu}+
\sqrt{\overline{\mu}^2-1-D\ell}}{(1+D\ell)^{1/2}}\Bigr)+
\frac{\pi^2}{6}\Theta^2
\sum_{\ell=0}^{\ell_{max}} f^\prime (\mu)\right\}
\end{eqnarray}
where $f^\prime (\mu)=\frac{\partial f_e}{\partial\overline{\varepsilon}_e^{\ell}}\mid_{
\overline{\varepsilon}_e^{\ell}=\mu}=\frac{\partial}{\partial\overline{\varepsilon}_e^{\ell}}
\sqrt{\overline{\varepsilon}_e^{\ell 2}-1-D\ell}\mid_{
\overline{\varepsilon}_e^{\ell}=\mu}$ and $\ell_{max}=\frac{\overline{\mu}^2-1}{D}$.

Taking into account the temperature correction, which follows from the last term of Eq.(\ref{ain}) in the limit $\mu\sim\hbar\omega_c\gg K_BT$, we write down the part of thermodynamic potential that depends on the temperature
\begin{eqnarray}
\label{tpart}
\Omega_T=-\frac{V m_ec^2 D}{12\lambda_c^3}\Bigl(\frac{K_BT}{m_ec^2}\Bigr)^2\left\{\frac{\overline{\mu}}{
\sqrt{\overline{\mu}^2-1}}+\frac{2\overline{\mu}\sqrt{\overline{\mu}^2-1-D}}{D}\right\}\ .
\end{eqnarray}
If the external magnetic field equals zero, $H_0=0$, then the expression (\ref{tpart}) reduces to
\begin{eqnarray}
\label{rex}
\Omega_T=-\frac{V m_ec^2 }{6\lambda_c^3}\Bigl(\frac{K_BT}{m_ec^2}\Bigr)^2\overline{\mu}\sqrt{\overline{\mu}^2-1} \ .
\end{eqnarray}
It should be noted that this equation is relativistic, from which follow expressions of the nonrelativistic limit \cite{landS}
\begin{eqnarray}
\label{nrex}
\Omega_T=-\frac{V K_B^2T^2\sqrt{2\varepsilon_F}m_e^{3/2} }{6\hbar^3}
\end{eqnarray}
and the ultrarelativistic limit \cite{landS}
\begin{eqnarray}
\label{urex}
\Omega_T=-\frac{V (\mu K_BT)^2}{6(\hbar c)^3}\ ,
\end{eqnarray}
where $\mu=cp_F$.

We now examine the ultrarelativistic limit, i.e. $\overline{\mu}=\frac{p_F}{m_ec}\gg 1$. Neglecting the temperature correction after integration over $\ell$ ($\sum_{\ell=1}^\infty\rightarrow\int_1^\infty d\ell$), we get
\begin{eqnarray}
\label{oo}
\Omega_0=-\frac{V m_ec^2 D}{4\pi^2\lambda_c^3}\overline{\mu}^2\left\{1+\frac{1}{3}\frac{\overline{\mu}^2}{
D}\Bigl(1-\frac{D}{\overline{\mu}^2}\Bigr)^{3/2}\right\}
\end{eqnarray}
and thus
\begin{eqnarray}
\label{pres}
P=-\frac{\Omega_0}{V}=\frac{ m_ec^2 D}{4\pi^2\lambda_c^3}\overline{\mu}^2\left\{1+\frac{1}{3}\frac{\overline{\mu}^2}{
D}\Bigl(1-\frac{D}{\overline{\mu}^2}\Bigr)^{3/2}\right\}\ .
\end{eqnarray}
Obviously, this equation in the absence of the magnetic field, i.e. $H_0=D=0$, reduces to the well known expression of the pressure
\begin{eqnarray}
\label{pres0}
P=\frac{1}{4}(3\pi^2)^{1/3}\hbar c n_e^{4/3} \ .
\end{eqnarray}
Expressions (\ref{oo}) and (\ref{pres}) describe cases, when $\overline{\mu}^2>D$ or $p_F^2>2\hbar eH_0/c$. In the opposite case, i.e. $\overline{\mu}^2<D$, we obtain
\begin{eqnarray}
\label{opp}
P=\frac{m_ec^2 D}{4\pi^2\lambda_c^3}\overline{\mu}^2\ .
\end{eqnarray}

Use of Eq.(\ref{nsmd}) in Eq.(\ref{opp}) yields the relation,
\begin{eqnarray}
\label{oirxuti}
P=\frac{m_ec^2\pi^2\lambda_c^3 }{D}\ n_e^2\ .
\end{eqnarray}
This formula shows that for given density $n_e$ the pressure of the electron gas decreases along with increase of the magnetic field as $P\sim\frac{1}{H}$.

If the frozen-in condition $\frac{n_e}{H}=const$ is satisfied, then Eq.(\ref{oirxuti}) casts as the equation of state of an ideal gas at the temperature $T=const$, i.e. $P\sim n_e$. This indicates that the magnetic field sufficiently changes the equation of state.

We next derive an expression for the thermodynamic potential or the pressure ($P=-\Omega/V$) from Eq.(\ref{ain}) at $T=0$ in the nonrelativistic limit $(p_F\ll m_ec $ and $\overline{\mu}=1+\frac{\varepsilon_F}{m_ec^2}$). The result is
\begin{eqnarray}
\label{oriexvsi}
P=\frac{(2m_e\varepsilon_F)^{3/2}\varepsilon_F}{3\pi^2\hbar^3}\left\{\eta+\frac{2}{5}(1-\eta)^{5/2}\right\} \ ,
\end{eqnarray}
where $\eta=\frac{\hbar\omega_c}{\varepsilon_F}.$

Equation (\ref{oriexvsi}) describes the thermodynamic properties in the nonrelativistic limit with arbitrary magnetic field, which for $H=0$ reduces to
\begin{eqnarray}
\label{orishvidi}
P=\frac{(3\pi^2)^{2/3}\hbar^2n_e^{5/3}}{5m_e}
\end{eqnarray}
and in the case, when $\eta >1$, reads
\begin{eqnarray}
\label{orirva}
P=\frac{(2m_e\varepsilon_F)^{3/2}}{3\pi^2\hbar^3}\ \varepsilon_F\eta=\frac{(2m_e\varepsilon_F)^{3/2}\hbar\omega_c}{
3\pi^2\hbar^3} \ .
\end{eqnarray}
To lucidly express the pressure by the magnetic field, employing Eq.(\ref{nsmd}) we rewrite Eq.(\ref{orirva}) as
\begin{eqnarray}
\label{oritsxra}
P=\zeta n_e\Bigr(\frac{n_e}{H}\Bigr)^2 \ ,
\end{eqnarray}
where $\zeta=\frac{\pi^4\hbar^4c^2}{3m_e e^2}$.
One can immediately see that there is a strong dependence between the pressure, the magnetic field and the density.
Noting the relation $E=-\frac{3}{2}\Omega$, we can also define the total energy as a function of the magnetic field.

With the thermodynamic potential $\Omega_T$ at hand from Eq.(\ref{tpart}), we now define the entropy and specific heat for the Fermi electron gas as
\begin{eqnarray}
\label{otsdaati}
S=-\Bigr(\frac{\partial\Omega}{\partial T}\Bigr)_{V,\mu} \hspace{1cm} and \hspace{1cm}
C_V=T\Bigr(\frac{\partial S}{\partial T}\Bigr)_{V,\mu} \ .
\end{eqnarray}

In the temperature limit $K_BT\ll\mu$ (for both cases: nonrelativistic, $T\ll\varepsilon_F/K_B=4.35\cdot 10^{-11}n_e^{2/3}deg$, and ultrarelativistic, $T\ll cp_F/K_B=9\cdot 10^{-1}n_e^{1/3}deg$) for the entropy we obtain
\begin{eqnarray}
\label{sami1}
S=\frac{VD}{6\lambda_c^3}\ \frac{K_BT}{m_ec^2}\ \overline{\mu}\left\{\frac{1}{\sqrt{\overline{\mu}^2-1}}+
\frac{2\sqrt{\overline{\mu}^2-1-D}}{D}\right\}
\end{eqnarray}
and the entropy per particle is
\begin{eqnarray*}
s=\frac{S}{N}=\frac{DK_BT\overline{\mu}}{6\lambda_c^3m_ec^2n_e}\ \left\{\frac{1}{\sqrt{\overline{\mu}^2-1}}+
\frac{2\sqrt{\overline{\mu}^2-1-D}}{D}\right\}\ .
\end{eqnarray*}

In the absence of external magnetic field for the entropy we get the relativistic expression
\begin{eqnarray}
\label{sami2}
s=\frac{K_BT\overline{\mu}\sqrt{\overline{\mu}^2-1}}{3\lambda_c^3n_em_ec^2}\ .
\end{eqnarray}

Obviously for the derivation of Eq.(\ref{sami1}) we assumed that $\overline{\mu}^2-1>D$. Now if the inequality $D>\overline{\mu}^2-1$ is valid for any value of the magnetic field, then for the entropy per electron we have
\begin{eqnarray}
\label{sami3}
s=\frac{D}{6\lambda_c^3m_ec^2}\ \frac{K_BT}{n_e}\ \frac{\overline{\mu}}{\sqrt{\overline{\mu}^2-1}} \ .
\end{eqnarray}

Since for an adiabatic process $S=const$, from Eqs.(\ref{sami1}), (\ref{sami2}) and (\ref{sami3}) we obtain
a new adiabatic equations. First, the adiabatic equation without the magnetic field reads
\begin{eqnarray}
\label{sami4}
\frac{K_BT}{n_e^{2/3}}\sqrt{1+\beta n_e^{2/3}}=const\ ,
\end{eqnarray}
where $\beta=\frac{(3\pi^2)^{2/3}\hbar^2}{m_e^2c^2}$.

From Eq.(\ref{sami4}) follow well known expressions for nonrelativistic limit, which is $T/n_e^{2/3}=const$, and for ultrarelativistic limit, which is $T/n_e^{1/3}$.

Next, a relativistic adiabatic equation for the case $ \overline{\mu}^2-1>D$ is derived from Eq.(\ref{sami1}) as
\begin{eqnarray}
\label{sami5}
\frac{H}{n_e}\frac{K_BT\sqrt{1+\frac{2\varepsilon_F}{m_ec^2}}}{\sqrt{\varepsilon_F}}\left\{
1+\frac{2\varepsilon_F}{\hbar\omega_c}\Bigl(1-\frac{\hbar\omega_c}{
\varepsilon_F}\Bigr)^{1/2}\right\}=const\ ,
\end{eqnarray}
where $\varepsilon_F=\frac{(2\pi^2)^{2/3}\hbar^2n_e^{2/3}}{2m_e\{\eta+\frac{2}{3}(1-\eta)^{3/2}\}^{2/3}}$.

Finally, in the case of rather strong magnetic field, i.e., $D>\overline{\mu}^2-1$ or $\hbar\omega_c>\varepsilon_F$, we obtain a novel adiabatic equation
\begin{eqnarray}
\label{sami6}
\frac{TH^2}{n_e^2}\sqrt{1+\beta\frac{n_e^2}{H^2}}=const \ ,
\end{eqnarray}
which yields in the nonrelativistic limit
\begin{eqnarray*}
\frac{TH^2}{n_e^2}=const
\end{eqnarray*}
and in the ultrarelativistic case
\begin{eqnarray*}
\frac{TH}{n_e}=const \ .
\end{eqnarray*}

We specifically note that the temperature T, as follows from Eq.(\ref{sami6}), is the function of volume and the magnetic field. The relation (\ref{sami6}) implies that at constant density the increase of the magnetic field consequently leads to the temperature decrease as $T\sim 1/H$ for ultrarelativistic case and
$T\sim 1/H^2$ for nonrelativistic case. It should be emphasized that, therefore, this is the adiabatic magnetization process for cooling the Fermi electron gas to ultra-low temperatures.

\section{SOUND VELOCITY}

We now determine the velocity of sound in a nonrelativistic degenerate Fermi gas and show a very  strong influence of the magnetic field on the expression of velocity. More importantly, we demonstrate that a new type of sound velocity arises, which we call the quantum magnetosound velocity, $C_s$, that is
\begin{eqnarray}
\label{sami7}
C_s^2=\frac{\partial (P,S)}{\partial (\rho,S)}=\Bigl(\frac{\partial P}{\partial \rho}\Bigr)_s \ ,
\end{eqnarray}
where $\rho=m_en_e$.

To calculate the velocity for various cases, we rewrite Eq.(\ref{sami7}) in a new variables $n_e$ and T. Namely,
\begin{eqnarray*}
C_s^2=\frac{\partial (P,S)}{\partial (n_e, T)}/\frac{\partial (\rho,S)}{\partial (n_e, T)} \ ,
\end{eqnarray*}
where the derivatives are written in Jacobian form.

First, in the case of absence of the magnetic field, from Eq.(\ref{sami7}) we get
\begin{eqnarray}
\label{sami8}
C_s^2=\frac{v_F^2}{3}+\frac{T^2}{6}\ \frac{m_ev_F}{\hbar^3 n_e}\ .
\end{eqnarray}

Next, we consider the case of a strong magnetic field, when the inequality $\hbar\omega_c>\varepsilon_F$ is satisfied. In this case, Eq.(\ref{ain}) yields the following expression for the pressure
\begin{eqnarray}
\label{sami9}
P=\hbar\omega_c\ n_e\left\{1+\frac{m_e^2T^2}{6\hbar^3n_ep_F}\right\}\ ,
\end{eqnarray}
where the second term is correction.

Having Eqs. (\ref{sami3}) and (\ref{sami9}), for the nonrelativistic limit $\overline{\mu}=1+\frac{\varepsilon_F}{m_e c^2}$ after a simple calculation we obtain the expression of the quantum magnetosound velocity as
\begin{eqnarray}
\label{otxi0}
C_s^2=\frac{\hbar\omega_c}{m_e}\left\{1+\frac{m_e^2T^2}{2\hbar^3n_ep_F}\right\}\ .
\end{eqnarray}

We specifically note here that this is a novel expression for the sound velocity due to the strong magnetic field.

\section{MAGNETISM OF FERMI GASES}

In this section, we take into account a spin of electron $\vec{s}$ in the expression of energy (\ref{eel}). The corresponding magnetic moment is $\vec{M}=N\cdot\vec{\mu}$, where N is the total number of electrons and $\vec{\mu}=\frac{\mu}{s}\ \vec{s}$ is the magnetic moment of electron, with $\mu=-\mu_B=-0.927\cdot 10^{-20}\frac{erg}{gauss}$.

With the expressions of thermodynamic potentials at hand, we can find the mean magnetic moment per unit volume of the Fermi gas as
\begin{eqnarray}
\label{otxi1}
\vec{M}=-\frac{1}{V}\Bigl(\frac{\partial \Omega}{\partial \vec{H}}\Bigr)_{T,V,\mu} \ .
\end{eqnarray}

We recall that, when $\hbar\omega_c>\varepsilon_F$, the thermodynamic potential has the form
\begin{eqnarray}
\label{otxi2}
\Omega_0=-\frac{V\pi^2\hbar^2c^2}{8e}\ \frac{n_e^2}{H} \ .
\end{eqnarray}

Use of this expression (\ref{otxi2}) in Eq.(\ref{otxi1}) yields the relation
\begin{eqnarray}
\label{otxi3}
\vec{M}=-\frac{\pi^2\hbar^2c^2}{8e}\ \frac{n_e^2}{H^2}\ \frac{\vec{H}}{H}=\chi\cdot\vec{H}\ .
\end{eqnarray}
Hence the susceptibility is
\begin{eqnarray}
\label{otxi4}
\chi=-\frac{\pi^2\hbar^2c^2}{8e}\ \frac{n_e^2}{H^3}\ .
\end{eqnarray}

It should be noted that the expression of susceptibility (\ref{otxi4}) indicates that in a strong magnetic field $H>\frac{cp_F^2}{e\hbar}$ the gas becomes diamagnetic and the transition to the superconducting state occurs. To show this noting Eq.(\ref{otxi4}), we write the magnetic field induction $\vec{B}$ in a Fermi electron gas
\begin{eqnarray}
\label{otxi5}
\vec{B}=\vec{H}+4\pi\vec{M}=(1+4\pi\chi)\vec{H} \ .
\end{eqnarray}

From Eqs.(\ref{otxi4}) and (\ref{otxi5}) one can immediately show the possibility of the existence of  Meissner's effect, i.e. $\vec{B}=0$ or $\chi=-\frac{1}{4\pi}$.

We estimate the magnetic field in which the metal goes to superconductor state ($4\pi\chi+1=0$) for the density $n_e\simeq 5\cdot 10^{22}1/cm^3$
\begin{eqnarray}
\label{otxi6}
H=\Bigl(\frac{\pi^3\hbar^2c^2}{2e}\ n_e^2\Bigr)^{1/3}\approx 10^7 \ gauss\ .
\end{eqnarray}

\section{OSCILLATION OF THERMODYNAMIC POTENTIAL}

As was mentioned above, in 1930 Landau has theoretically predicted that the thermodynamic potential or the partition function contains an oscillatory terms in the variable $2\mu/\hbar\omega_c=\mu/\beta_B H$. The same year the periodic behavior of susceptibility of metals at low temperatures and in a strong magnetic fields was experimentally discovered by de Haas-van Alphen.

In order to investigate the oscillatory part of thermodynamic potential, we recall Eq.(\ref{tpot}). Up to now we have evaluated Eq.(\ref{tpot}) only in the case when $\varepsilon_F\sim\hbar\omega_c\gg K_BT$. In this case the oscillatory part of Eq.(\ref{tpot}) is very small in comparison with a smooth part. We now consider the case of the magnetic field in the range $\varepsilon_F\gg\hbar\omega_c\geq K_BT$. These inequalities imply that the non-oscillatory part of the thermodynamic potential practically does not depend on the magnetic field and we, therefore, can use the above equations at H=0.

For the oscillatory part of the thermodynamic potential we write the last part of Eq.(\ref{pois}) as
\begin{eqnarray}
\label{otxi7}
\tilde{\Omega}=\frac{Vm_eT}{\pi^2\hbar^3}Re
\sum_{k=1}^\infty\int_0^\infty d\ell\ \Psi(\ell)e^{\imath 2\pi k\ell}\ ,
\end{eqnarray}
where
\begin{eqnarray*}
\Psi(\ell)=-\hbar\omega_c\ \int_{-\infty}^\infty dp_z\ln
\Bigl(1+e^{\frac{\mu-\varepsilon_e^\ell}{T}}\Bigr) \ .
\end{eqnarray*}

We can rewrite the integral in Eq.(\ref{otxi7}) in dimensional variables in the form
\begin{eqnarray*}
\int_0^\infty d\ell\ \Psi(\ell)e^{\imath 2\pi k\ell}=-\hbar\omega_cm_ec\ \int_{-\infty}^\infty du
\int_0^\infty d\ell\ln\Bigl(1+\frac{\overline{\mu}-\overline{\varepsilon}_e^\ell}{\Theta}\Bigr)e^{\imath 2\pi k\ell}\ .
\end{eqnarray*}

If we now replace the variable $\ell$ by $\overline{\varepsilon}_e^\ell$ in these integrals, we obtain
\begin{eqnarray}
\label{otxi8}
\int_0^\infty d\ell\ \Psi(\ell)e^{\imath 2\pi k\ell}=-\frac{\hbar\omega_cm_ec}{D}\ 2\int_{-\infty}^\infty du
\ e^{-\imath\alpha u^2}\ \int_{\sqrt{1+u^2}}^\infty d\overline{\varepsilon}_e^\ell\ \overline{\varepsilon}_e^\ell\ln\Bigl(1+e^{\frac{\overline{\mu}-\overline{\varepsilon}_e^\ell}{\Theta}}\Bigr)
\ e^{\imath 2\pi k\frac{\overline{\varepsilon}_e^{\ell 2}-1}{D}}\ ,
\end{eqnarray}
where $\alpha=\frac{2\pi k}{D}$.

First, we integrate the second integral in Eq.(\ref{otxi8}). To this end, we employ a new variable $\overline{\varepsilon}_e^\ell-\overline{\mu}=\Theta z$, so that we rewrite the second integral as
\begin{eqnarray*}
A=e^{-\imath\alpha}\ \Theta\ \int_{\frac{\sqrt{1+u^2}-\overline{\mu}}{\Theta}}^\infty dz\ (\overline{\mu}+\Theta z)\ \ln(1+e^{-z})\ e^{\imath 2\pi k\frac{\overline{\mu}^2+2\overline{\mu}\Theta z+\Theta^2z^2}{D}}\ .
\end{eqnarray*}
Since $\overline{\mu}\gg\Theta$, the lower limit $\frac{\sqrt{1+u^2}-\overline{\mu}}{\Theta}\rightarrow -\infty$. Thus, in the exponent we can neglect $\Theta z$ in comparison with $\overline{\mu}$ and therefore
\begin{eqnarray*}
A=\Theta\ e^{\imath\alpha(\overline{\mu}^2-1)}\ \overline{\mu}\ \int_{-\infty}^\infty dz\ \ln(1+e^{-z})\ e^{\imath\frac{4\pi\overline{\mu}\Theta}{D}kz}\ .
\end{eqnarray*}
After integration twice by parts, we finally obtain the required result
\begin{eqnarray*}
A=\Theta\ e^{\imath\alpha(\overline{\mu}^2-1)}\ \overline{\mu}\ \frac{1}{\beta^2}\int_{-\infty}^\infty dz
\ \frac{e^z}{(e^z+1)^2}\ e^{\imath\beta z}=\Theta \overline{\mu}\ e^{\imath\alpha(\overline{\mu}^2-1)}\ \frac{\pi\beta}{\sinh\pi\beta}\ ,
\end{eqnarray*}
where $\beta=\frac{4\pi\overline{\mu}\Theta k}{D}$. The advantage of this calculation is that the integrand in the second integral of Eq.(\ref{otxi8}) contributes practically only in the vicinity of the Fermi energy.

The integration over u in Eq.(\ref{otxi8}) is simple, and gives
\begin{eqnarray}
\label{otxi9}
\int_{-\infty}^\infty du\ e^{-\imath\alpha u^2}=\sqrt{\frac{\pi}{\alpha}}e^{-\imath\pi/4} \ .
\end{eqnarray}

Taking into account Eqs. (\ref{otxi7}), (\ref{otxi8}) and (\ref{otxi9}), finally we have for the oscillatory part of thermodynamic potential
\begin{eqnarray}
\label{xuti0}
\tilde{\Omega}=\frac{V(m_e\hbar\omega_c)^{3/2}}{2\pi^2\hbar^3}\ K_BT\ \sum_{k=1}^\infty\ \frac{1}{k^{3/2}}\ \frac{\cos [\frac{\pi km_ec^2}{\hbar\omega_c}(\overline{\mu}^2-1)-\frac{\pi}{4}]}{\sinh(2\pi^2k\overline{\mu}K_BT/\hbar\omega_c)}\ .
\end{eqnarray}
This is the relativistic expression of $\tilde{\Omega}$ and indeed contains oscillatory terms in the variable $\frac{m_ec^2}{\hbar\omega_c}(\overline{\mu}^2-1)$.

In the nonrelativistic limit ($\overline{\mu}^2-1=\varepsilon_F/m_ec^2$), from Eq.(\ref{xuti0}) follows a well known expression of $\tilde{\Omega}$ \cite{landS}. Whereas, in the ultrarelativistic case ($\overline{\mu}=\frac{p_F}{m_ec}\gg 1$), for the cosine in Eq.(\ref{xuti0}) we have
\begin{eqnarray*}
\cos [\frac{\pi p_F^2}{m_e\hbar\omega_c}\ k-\frac{\pi}{4}]\ .
\end{eqnarray*}

We now calculate the magnetic moment as the derivative of Eq.(\ref{xuti0}). To this end, we differentiate cosine alone to obtain
\begin{eqnarray}
\label{xuti1}
M=-\beta_BV\ \frac{m_e^{3/2}K_BTm_ec^2(\overline{\mu}^2-1)}{\pi\hbar^2(\hbar\omega_c)^{1/2}}
\sum_{k=1}^\infty\ \frac{1}{\sqrt{k}}\frac{\sin[\frac{\pi k}{\hbar\omega_c}m_ec^2(\overline{\mu}^2-1)-\frac{\pi}{4}]}{\sinh(2\pi^2k\overline{\mu}K_BT/\hbar\omega_c)}\ ,
\end{eqnarray}
and hence the susceptibility is
\begin{eqnarray}
\label{xuti2}
\chi_{osc}=\frac{M}{H}=-\frac{VK_BT\beta_B^2m_e^{3/2}m_ec^2(\overline{\mu}^2-1)}{\pi\hbar^2(\hbar\omega_c)^{3/2}}
\sum_{k=1}^\infty\ \frac{1}{\sqrt{k}}\frac{\sin[\frac{\pi km_ec^2(\overline{\mu}^2-1)}{\hbar\omega_c}-\frac{\pi}{4}]}{\sinh(2\pi^2k\overline{\mu}K_BT/\hbar\omega_c)}\ .
\end{eqnarray}

As one can see the susceptibility has an oscillatory contribution in the variable
\begin{eqnarray*}
\frac{m_ec^2(\overline{\mu}^2-1)}{\hbar\omega_c}\gg 1 \hspace{.5cm} or \hspace{.5cm}
\frac{m_ec^2(\overline{\mu}^2-1)}{\hbar}\gg \omega_c\ ,
\end{eqnarray*}
which means that the susceptibility oscillates with a high frequency. We note here that for both nonrelativistic and relativistic cases these inequalities are satisfied.

We also note that since we consider the case of $\mu\gg\hbar\omega_c\sim K_BT$, the smooth part of the magnetic moment is smaller than the oscillatory one as $[\hbar\omega_c/m_ec^2(\overline{\mu}^2-1)]^{1/2}$.

We now derive the expression for $\tilde{\Omega}$ at low temperature $K_BT\ll\hbar\omega_c$, and $k=1$, i.e. $\sinh x=x(1+\frac{x^2}{6})$, where $x=\frac{2\pi^2\overline{\mu}K_BT}{\hbar\omega_c}\ll 1$. The result is
\begin{eqnarray}
\label{xuti3}
\tilde{\Omega}=\frac{V m_e^{3/2}(\hbar\omega_c)^{5/2}}{(2\pi^2)^2\hbar^3\overline{\mu}n_e}[1-\frac{
(2\pi^2\overline{\mu})^2(K_BT)^2}{6(\hbar\omega_c)^2}]\cos[\frac{\pi m_ec^2(\overline{\mu}^2-1)}{\hbar\omega_c}-\frac{\pi}{4}] \ .
\end{eqnarray}

Having the expression for the oscillatory part of thermodynamic potential (\ref{xuti3}), we can calculate the entropy, specific heat and the sound velocities. Namely, for the entropy we get
\begin{eqnarray}
\label{xuti4}
\tilde{S}=-\Bigl(\frac{\partial \tilde{\Omega}}{\partial T}\Bigr)_{V,\mu}=\frac{V m_e^{3/2}\overline{\mu}}{
3\hbar^3n_e}(\hbar\omega_c)^{1/2}K_BT\cos[\frac{\pi m_ec^2(\overline{\mu}^2-1)}{\hbar\omega_c}-\frac{\pi}{4}] \ .
\end{eqnarray}
Next, for the specific heat we have
\begin{eqnarray}
\label{xuti5}
\tilde{C}_V=T\Bigl(\frac{\partial \tilde{S}}{\partial T}\Bigr)_{V,\mu}=\tilde{S}\ .
\end{eqnarray}
Finally, in the nonrelativistic limit ($\overline{\mu}^2-1=\frac{p_F^2}{m_e^2c^2}\ll 1$), for the adiabatic sound total velocity we obtain
\begin{eqnarray}
\label{xuti6}
C_S^2=\frac{1}{m_e}\Bigl(\frac{\partial P}{\partial n_e}\Bigr)_S=\frac{v_F^2}{3}+\frac{(\hbar\omega_c)^{3/2}}{2\pi(3\pi^2n_e)^{1/3}\hbar m_e^{1/2}}\sin[\frac{\pi
p_F^2}{m_e\hbar\omega_c}-\frac{\pi}{4}] \ .
\end{eqnarray}

Clearly, as the susceptibility, the entropy, the specific heat and the sound velocity oscillate with a high frequency.

\section{SUMMARY}

We have created the relativistic thermodynamics of magnetized quantum electron gas. To this end, we have derived the thermodynamic potential of relativistic electron gas with an arbitrary magnetic field and temperature, and investigated all the thermodynamic quantities as a function of density, temperature and the magnetic field. In particular cases, as the non-relativistic and the ultra-relativistic,
expressions of the entropy, the specific heat, the pressure and the total energy are explicitly found. We have shown that the magnetic field sufficiently changes the equation of state. We have obtained a novel set of adiabatic equations. A novel adiabatic equation implies that at the constant density the increase of the magnetic field consequently leads to the temperature decrease as $T\sim 1/H$ for ultrarelativistic case and
$T\sim 1/H^2$ for nonrelativistic case. Therefore, we are here suggesting the adiabatic magnetization process for cooling the Fermi electron gas to ultra-low temperatures as an alternative to the known adiabatic demagnetization mechanism. We have also obtained expressions of sound velocity for several cases and demonstrated a very strong influence of the magnetic field. Moreover, we have revealed a new type of sound velocity, which we have named as the quantum magnetosound velocity. Finally, we have shown that in a strong magnetic field the Fermi gas becomes diamagnetic and the transition to the superconducting state takes place. The study of thermodynamics of a Fermi gas of sufficiently low temperature and in a strong magnetic field is of fundamental significance for understanding physics of compact astrophysical objects, e.g., the interior of white dwarf stars, magnetospheres of neutron stars and magnetars. The results of the present paper may be of substantial interest in connection with the applications in modern technology (e.g. metallic and semiconductor nanostructures), new kinds of magnetic refrigerator designs, next generation intense laser-solid density matter experiments.

\end{document}